\documentclass[graphicx,amsmath,amssymb,pra,twocolumn]{revtex4-1}

\usepackage{graphicx}		
\usepackage{color}
\usepackage{amsmath}
\usepackage{subfigure}
\usepackage{upgreek}
\usepackage{footmisc}

\usepackage{dsfont}

\begin{document}
	
	\title{Noise-induced network topologies}	
	\author{Frederic Folz$^1$}
	\author{Kurt Mehlhorn$^2$}
	\author{Giovanna Morigi$^1$}
	\affiliation{$^1$Theoretische Physik, Universit\"at des Saarlandes, 66123 Saarbr\"ucken, Germany \\ $^2$Algorithms and Complexity Group, Max-Planck-Institut f\"ur Informatik, Saarland Informatics Campus, 66123 Saarbr\"ucken, Germany}
	
	\date{\today}
	
	\begin{abstract} 
We analyze transport on a graph with multiple constraints and where the weight of the edges connecting the nodes is a dynamical variable. The network dynamics results from the interplay between a nonlinear function of the flow, dissipation, and Gaussian, additive noise. For a given set of parameters and finite noise amplitudes, the network self-organizes into one of several meta-stable configurations, according to a probability distribution that depends on the noise amplitude $\alpha$. At a finite value $\alpha$, we find a resonant-like behavior for which one network topology is the most probable stationary state. This specific topology maximizes the robustness and transport efficiency, it is reached with the maximal convergence rate, and it is not found by the noiseless dynamics. We argue that this behavior is a manifestation of noise-induced resonances in network self-organization. Our findings show that stochastic dynamics can boost transport on a nonlinear network and, further, suggest a change of paradigm about the role of noise in optimization algorithms. 
\end{abstract}
	
\maketitle
	
The ability to extract information from large data bases has become essential to modern science and technologies. This quest is central to foundational studies, such as in astronomy, for shedding light on the constitution of our universe \cite{Sen:2022}, and in particle physics, for efficiently identifying relevant events in high-energy physics experiments \cite{Coelho:2021}, as well as to applications, such as the design of efficient power grids \cite{Kezunovic:2020} and the sustainable exploitation of water supplies \cite{Gohil:2021}. A question lying at the core of these efforts is: What are the key ingredients and dynamics at the basis of an efficient search in a generic database? This question encompasses a large number of physically relevant situations, including the determination of the ground state of a quantum many-body problem \cite{Santoro:2002,Carleo:2017,Tilly:2022}, the transport of excitons \cite{Huelga:2021,Mattiotti:2022} and cells \cite{Shaebani:2020,Meyer:2021}, and the search for food by living organisms \cite{Gao:2019,Meyer:2017b}. The latter is a precious source of insights because of organisms' capability to extract information from and adapt to a dynamically changing environment \cite{Gao:2019,Yang:2020}. One example is the food search of Physarum polycephalum and of ant colonies, that have inspired optimization algorithms successfully applied to real-world optimization problems \cite{Tero:2007,Meyer:2017b,Gao:2019,Li:2020,Ornek:2022}.

One relevant aspect of biological systems is the capability to efficiently extract relevant information for their survival in a noisy environment, where parameters fluctuate and the amount and location of food sources can change over time. For instance, models simulating excitable systems, such as forest fires \cite{Meron:1992} and neurons \cite{Lindner:2004}, show that noise can lead to qualitatively different effects. These include phenomena such as stochastic and coherence resonance \cite{Gammaitoni:1998,Lindner:2004,Perc:2007}, synchronization \cite{Nakao:2007,Boccaletti:2002}, and noise-induced phase transitions \cite{VandenBroeck:1994,Sagues:2007}. 
A systematic understanding of the role of noise in a search problem would shed light on its role in cooperative dynamics, including neural networks, and might initiate novel applications to optimization problems. 

\begin{figure*}
		\includegraphics[width=0.7\textwidth]{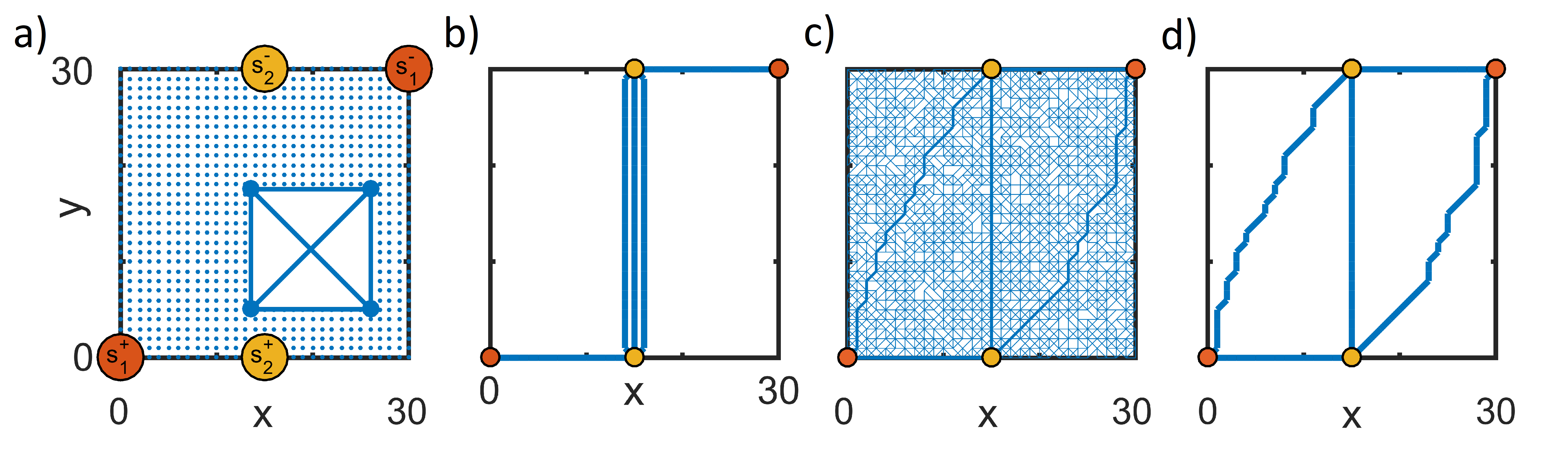}
		\caption{\label{Fig:1} (a) Network self-organization is simulated on a grid of 31 x 31 nodes with two demands. The demands are indicated by the pairs of red and yellow nodes, the sources are labelled by $s_+^i$, the sinks by $s_-^i$, the inset shows that the nodes are connected by horizontal, vertical, and diagonal edges. The network design results from the dynamics of the edges, which are modelled by time-varying conductivity $D_{u,v}$ on an electrical network and in the presence of additive noise according to Eq.\ \eqref{Eq:Q} and \eqref{Eq:D}. (b) and (c) display the networks reached after a sufficiently long integration time in the noiseless case ($\alpha=0$) and for $\alpha=0.002$, respectively. The widths of the edges are proportional to the corresponding amplitude of $D_{u,v}$. (d) displays the multi-scale backbone extracted from (c) using a filtering procedure (see text). See Fig.\ \ref{Fig:2} for details on the numerical simulations.}
\end{figure*}

In this work, we analyze the self-organization dynamics of a network in the presence of additive noise and with multiple constraints to be satisfied. The constraints are two pairs of source and sink nodes, as illustrated in Fig.\ \ref{Fig:1}(a), at which a constant flow is injected and extracted, respectively. In computer science, it is a multi-commodity problem: each pair of source and sink is a {\it demand} to be satisfied and the path satisfying the demand is a {\it flow of commodity} \cite{Bonifaci:2022,Lonardi:2022}. Examples are a city transport network, where each commodity is the passengers travelling between two stations, or an electrical circuit, where the commodity is the electrical current satisfying a given potential difference between two nodes. The optimal path is a network topology obtained by integrating a set of equations for the graph's nodes and edges, where the strength of the edges, determining the edge capacity \cite{Bonifaci:2022}, is a dynamical variable subject to the competition between dissipation and an activation force depending on the total flow across the edge \cite{Tero:2007,Gao:2019,Bonifaci:2022,Lonardi:2022}. In the absence of noise, the dynamics tends to identify the optimal path satisfying the constraints according to a rule that promotes transport along shared routes and instead inhibits it when the flow along one edge is below a chosen threshold. Differing from the typical settings, in this work, we assume that the edge capacity can also fluctuate due to a Langevin force \cite{vanKampen:2007}. We show that the introduction of stochasticity has a dramatic impact on the convergence to the optimal path. Among several noteworthy features, the solutions follow a multi-stable distribution that undergoes discontinuous transitions as a function of the noise amplitude. Remarkably, the distribution exhibits a resonant-type of behavior as a function of the noise strength.
In fact,  for a finite range of noise amplitudes the network self-organizes into a topology that maximizes its robustness and that is not found by the noiseless dynamics. 

{\it The model.} In the following, we will refer to the multi-commodity problem in terms of currents in an electrical circuit, keeping in mind that this is just one possible example. The edge capacity is determined by the conductivity, which is a  dynamical variable. The circuit consists of a spatial grid composed of $31 \times 31$ nodes. Each node, labelled $u$, can connect to a number of nearest and next-nearest neighbors, described by the set $E_u$ (see the inset of Fig.\ \ref{Fig:1}(a)). 
The emerging networks need to serve two demands $i=1,2$, each represented by a source node $s_+^i$ and a sink node $s_-^i$, where a current is injected ($+I_i$) and extracted ($-I_i$), respectively. Each demand generates a flow across the network: The flow of the demand $i$ is composed of the contributions $Q_{u, v}^i$ at the edge connecting nodes $(u,v)$. The flow of each demand is conserved at each node $u$, $\sum_{v \in E_u} Q_{u, v}^i=0$ (Kirchhoff's law) except for the source and sink where $\sum_{v \in E_{s_i^\pm}} Q_{s_i^\pm, v}^i=\pm I_i$. The flow of the demand $i$ along the edge $(u,v)$ is proportional to the edge conductivity $D_{u, v}(t)$ and to the difference between the potentials of the two nodes $p_u^i(t)$ and $p_v^i(t)$:
\begin{equation}
\label{Eq:Q}
    Q_{u, v}^i(t) = \frac{D_{u, v}(t)}{L_{u, v}} (p_u^i(t) - p_v^i(t))\,,
\end{equation}
where $L_{u, v}$ is the edge length and is constant. The edge dynamics is described by the coupled dynamical variables $p_u^i$ and $D_{u, v}$. The potential $p_u^i$ is determined for each demand $i$ as a function of $D_{u, v}(t)$ by solving the linear set of equations in Eq.\ \eqref{Eq:Q} with Kirchhoff's law, as detailed in Ref.\ \cite{Bonifaci:2022} and in the Supplemental Material (SM) \footnote{\label{SM} See Supplemental Material at [URL will be inserted by publisher] for (1) the parameter values that we used, details on numerical simulations and details on calculating the potential $p_u^i$ at node $u$, (2) the steady state and the convergence rate, (3) the disparity filter, (4) the robustness of the network, (5) the dependence on the injection current and for a larger number of demands and (6) movies of the dynamics leading to the networks (A)-(G).}. The conductivity $D_{u, v}(t)$ obeys the stochastic nonlinear equation \cite{Meyer:2017,Folz:2021}:
\begin{equation}
\label{Eq:D}
    \partial_t D_{u, v} = f(Q_{u,v}) - \gamma D_{u, v}+\sqrt{2\gamma}\alpha\xi_{u, v}(t)\,.
\end{equation}
Here, $f(x)$ is the activation function with sigmoidal form:  $f(x)=x^n/(\kappa^n + x^n)$ with $n>0$ (in what follows we choose $n=1.2$), the argument is the total flow along the edge, $Q_{u, v}=\sum_i|Q_{u, v}^i|$, and $f$ saturates when $Q_{u, v}$ exceeds the threshold $\kappa$. Hence, $f(x)$ gives rise to an effective interaction between demands that favors the sharing of transport routes between commodities. The activation is counteracted by dissipation at rate $\gamma$. Fluctuations in the conductivity are simulated by the stochastic force $\xi(t)$, whose amplitude is scaled by the parameter $\alpha$. The force is statistically defined by the average over an ensemble of trajectories: it has no net drift, $\langle \xi_{u, v} (t) \rangle = 0 $, and simulates Gaussian white noise, $\langle \xi_{u, v}(t) \xi_{u', v'}(t') \rangle = \delta_{u, u'} \delta_{v, v'} \delta(t - t')$ \cite{vanKampen:2007} \footnote{We note that the variable $\alpha$ here is physically equivalent to the temperature $T$ of an external bath according to the relation $T\propto \alpha^2$ \cite{vanKampen:2007}.}.

\begin{figure*}
		\includegraphics[width=1\textwidth]{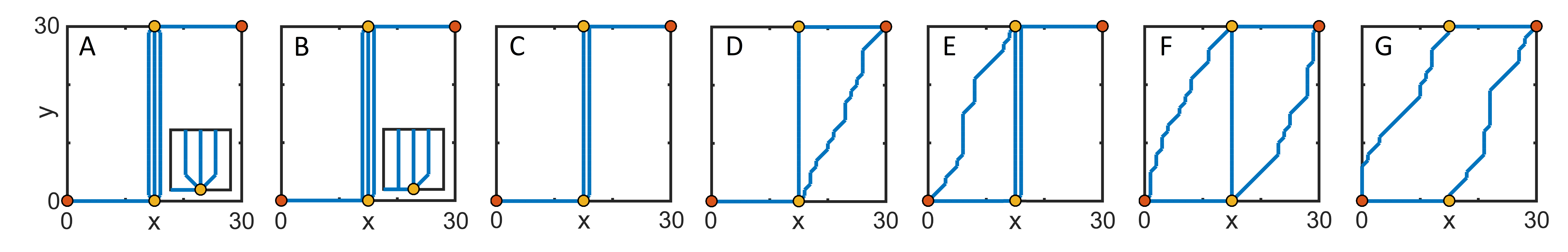}
		\caption{\label{Fig:2} Network topologies for increasing values of the noise amplitude $\alpha$ (from $\alpha=0$ to $\alpha=0.005$). (A) is the noiseless case, (B)-(G) are the typical backbones for $\alpha>0$, the probability of their occurrence depends on $\alpha$ and is shown in Fig. \ref{Fig:3} (for (D),(E) we report one of the two symmetric configurations). The networks are the result of the time evolution of Eqs.\ \eqref{Eq:Q} - \eqref{Eq:D} for a time $t=250 \gamma^{-1}$ imposing  $I_1=I_2= 0.45$ and $\kappa = 1$ . Initially, we set $D_{u,v}=0.5$ on all edges. The integration of Eq.\ (2) is performed using the Euler-Maruyama scheme \cite{Kloeden_1992} with step size $\Delta t = 0.1 \gamma^{-1}$. In the SM [29] movies are reported which show how the dynamics at different noise amplitudes leads to each of the topologies. 
		}   
\end{figure*}

Our model shares analogies with resistor networks \cite{Kaiser:2021} but is essentially different in that the edge conductivities (the metric) are dynamical variables. Equations \eqref{Eq:Q} and \eqref{Eq:D}, in the absence of noise, were used in Ref.\ \cite{Tero:2007} for modelling the structures built by a uni-cellular organism for food search in a maze \cite{Nakagaki:2000} and on a graph simulating the Tokyo railroad system \cite{Tero:2010}. These equations set the basis for optimization algorithms \cite{Gao:2019} and have been applied to multi-commodity problems \cite{Bonifaci:2022,Lonardi:2022} using other classes of activation functions than the sigmoidal functions. The studies of Ref.\ \cite{Bonifaci:2022,Lonardi:2022} showed that the dynamics converges towards networks optimizing between the sharing of transport routes, favored by the activation function, and the total cost of the network (here given by the total length of the edges of the closed paths) that is controlled by dissipation. In Refs.\ \cite{Meyer:2017,Folz:2021}, stochastic forces were added to the model for one single demand connected by two paths of the same length but different, periodically varying, dissipation rates. In \cite{Folz:2021}, the resulting flow was analyzed as a function of the frequency of the dissipation rates and amplitude of the noise, manifesting the characteristic features of stochastic resonance and noise-induced limit cycles. In this work, we analyze, for the first time, a multi-commodity problem in the presence of noise. 
The relatively simple geometry of our problem allows us to single out the essential features and visualize the manifold of topologies as a function of the noise amplitude. 

{\it Results.} We integrate Eqs.\ \eqref{Eq:Q} and \eqref{Eq:D} with the static boundary conditions of Fig.\ \ref{Fig:1}(a) after initializing the conductivities on all edges to the same value (see also SM \footnotemark[1]). The system evolution thus initially consists of redirecting the flow along edges by modifying the conductivities. For $\alpha=0$, the dynamics is noiseless and converges to the configuration of Fig.\ \ref{Fig:1}(b): the flow satisfying both demands is routed along the vertical connection. The system tends to generate parallel routes. In fact, the transport along one edge is bound to a maximal value due to the saturation of the sigmoidal function. For $\alpha>0$, we integrate stochastic differential equations. Figure \ref{Fig:1}(c) displays a network configuration obtained by integrating the stochastic dynamics for one trajectory and after a sufficiently long simulation time. It is evident that noise leads to a fluctuating distribution of weak connections. In order to be able to perform a classification, we apply a filter mechanism to each trajectory as follows.  We level out the fluctuations by taking the time average of the configurations in the regime where the simulation has converged. We then account for the statistical relevance of the links by means of the disparity filter of Ref.\ \cite{Serrano:2009} (see SM \footnotemark[1]). Figure \ref{Fig:1}(d) displays the network topology extracted from (c) after applying the disparity filter to the time-averaged configuration. For each value of $\alpha$, we evaluate 5000 trajectories. 


Figure \ref{Fig:2} shows the typical network topologies ordered by increasing noise amplitude, starting from the noiseless case (A). Each is unique in terms of connectivity of the hubs and is characterized by a different set of values of the measures we apply, as we detail later. The networks (B),(C) are found for small $\alpha > 0$ and are similar to the noiseless case with the tendency to decrease the shared routes. In addition, (C) decreases the number of connections. Configurations (B)-(E) are multi-stable and generally break the point symmetry of the configuration. For larger values of $\alpha$, the topologies converge to one of the two configurations (F),(G), with a bi-stable region about $\alpha\sim 3\times 10^{-3}$. Topologies (F),(G) are point symmetric but qualitatively different from (A). Note that (A)-(G) are fixed points of the noiseless dynamics. Noise dramatically modifies the respective basin of attraction as visible by analysing the network measures as a function of $\alpha$. 

\begin{figure*}
		\includegraphics[width=1\textwidth]{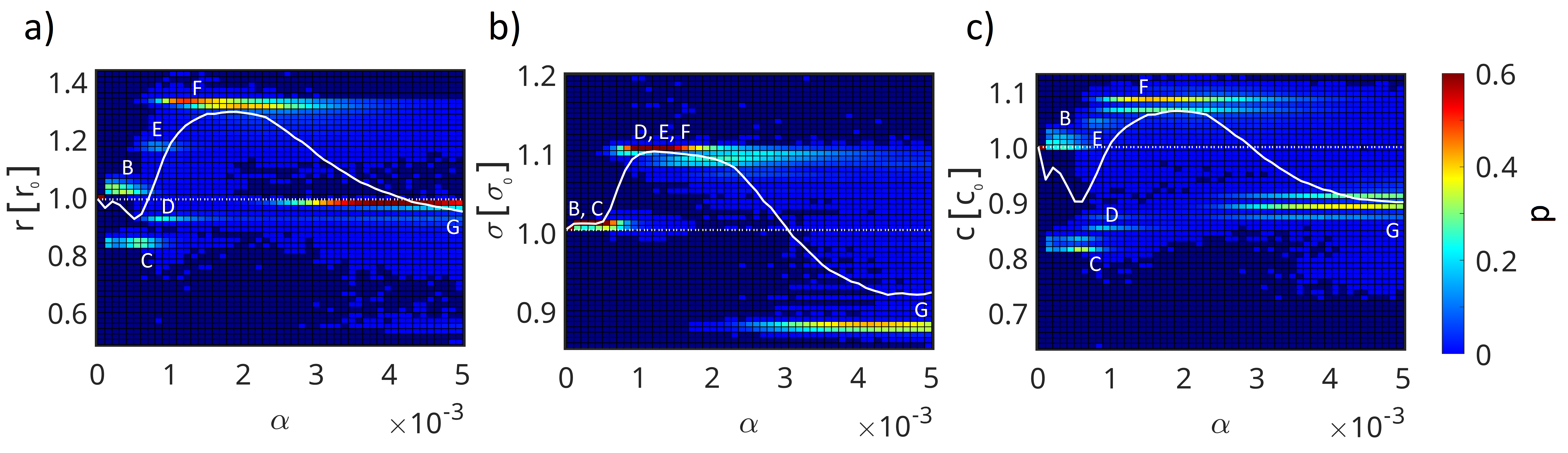}
		\caption{\label{Fig:3} Network measures as a function of the noise amplitude $\alpha$: (a) robustness $r$, (b) transport efficiency $\sigma$, and (c) network cost $c$. Each measure is in units of the respective value $r_0$, $\sigma_0$, and $c_0$ for $\alpha = 0$ (dashed line in the plot). The white solid line is the mean value taken over 5000 trajectories at each value of $\alpha$, the color scale gives the fraction of trajectories for each value of $r$, $\sigma$, $c$: dark blue is statistically irrelevant, dark red corresponds to 60\%. The distribution clusters about a set of the topologies (the labels follow the legend of Fig.\ \ref{Fig:2}) and undergoes discontinuous transitions as $\alpha$ is varied. For $\alpha\in [0.001-0.003]$, it narrows about a single topology, (F), with optimal robustness and transport efficiency.}
\end{figure*}		

The network measures are determined on the backbone of each trajectory: (i) The robustness $r$ provides information on the quality of the connections: it increases by adding paths connecting two nodes, which in turn makes the network more robust against edge failures. It is defined by $ r = 1/(\sum_{i=1}^2 R_i/2 )$, with  $R_i = (p_{s_+^i}^i - p_{s_-^i}^i)/I_i$ as the effective resistance between the source node $s_+^i$ and the sink node $s_-^i$ of each demand $i$, see Ref.\ \cite{Ellens:2013} and the SM \footnotemark[1]. (ii) The transport efficiency $\sigma$ is given by ${1/\sigma =\sum_{i=1}^2 d_i/2}$, where $d_i$ is the length of the shortest path connecting $s_i^+$ and $s_i^-$ \cite{Tero:2010}. (iii) Finally, the cost of the network $c$ is the total length, found by summing over the ensemble $\mathcal{E}$ of segments $L_{u, v}$ of the backbone where the conductivity is non-zero \cite{Tero:2010}, 
$ c= \sum_{(u, v) \in \mathcal{E}} L_{u, v}$.
The measures $(r,\sigma,c)$ are displayed in Fig.\ \ref{Fig:3}(a)-(c) as a function of the noise amplitude $\alpha$. The white lines indicate their mean values. The slope of the mean robustness and cost at $\alpha=0$ is negative, showing that - on average - for small noise amplitudes the dynamics converges to topologies with worse robustness and lower cost than for the noiseless case. After this transient, they all reach a maximum for an interval of noise amplitudes centered about $\alpha \sim 2 \times 10^{-3}$ that is qualitatively above the noiseless value. For each value of $\alpha$ the distribution of $x=r,\sigma,c$ about the mean is encoded in the color scale. The distribution is clustered about the topologies of Fig. \ref{Fig:2} with probabilities depending on $\alpha$.
One striking feature is that (A) disappears for $\alpha>0$, indicating that it is unstable against fluctuations. As $\alpha$ is increased, the system jumps to different configurations, undergoing discontinuous, noise-induced transitions. The topologies (B)-(E) occur at low, non-vanishing values of $\alpha$ and are generally multi-stable. Remarkably, for a non-zero interval of values $\alpha$ (in the range $0.001-0.003$) the distribution narrows and becomes single-peaked and the dynamics converges to (F). This topology optimizes both robustness and transport efficiency, with a qualitative improvement over (A). At even larger amplitudes $\alpha$, first (F) coexists with (G), then (G) becomes the most probable configuration. Network (G) has the same robustness as (A). Its worse transport efficiency and lower cost are due to noise: the number of statistically relevant edges decreases with $\alpha$. The distribution about (G) is broader according to the common expectation that noise increases the variance. Instead, the narrowing at $\alpha\in[0.001-0.003]$ about the topology (F) contradicts this intuition. 

\begin{figure}
		\includegraphics[width=0.4\textwidth]{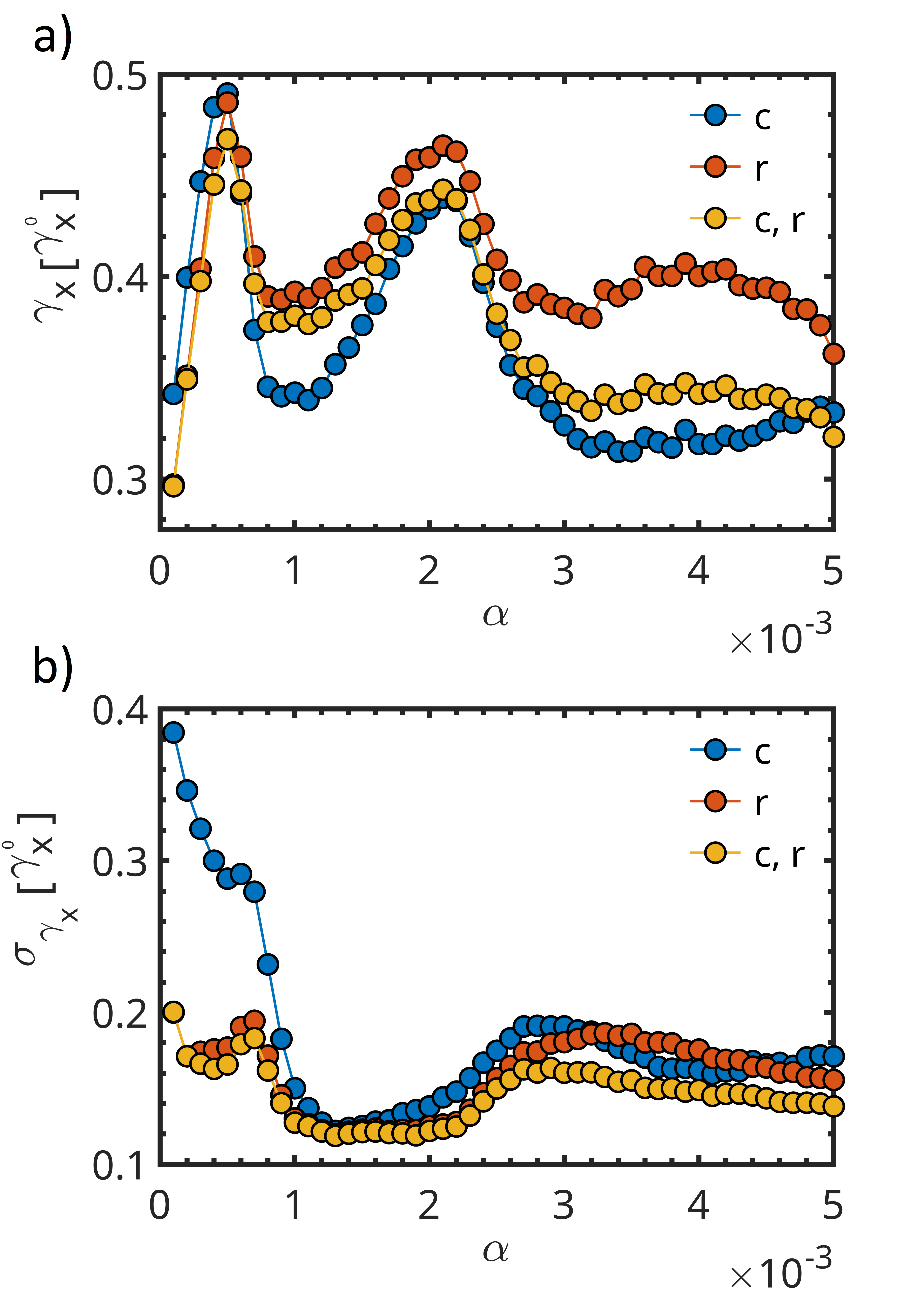}
		\caption{\label{Fig:4} {(a) Average convergence rate $\gamma_x$ as a function of the noise amplitude $\alpha$ (in units of the respective value $\gamma_x^0$ for the noiseless case). $\gamma_x$ is the inverse of the time that a trajectory needs to reach a stationary value of the cost (blue), of the robustness (red), and of joint cost and robustness (yellow), see SM for the definition [29]. (b)  displays the corresponding variance $\sigma_{\gamma_x}$ (in units of $\gamma_x^0$). The averages are taken over an ensemble of 5000 trajectories. About $\alpha\sim 0.002$, the dynamics converges to (F).}} 	
\end{figure}

The trajectories converge relatively fast towards one of the topologies of Fig.\ \ref{Fig:2}. Figure \ref{Fig:4}(a) displays the average convergence rates to a stationary value of $r,\sigma,c$ as a function of $\alpha$.  The rates are not monotonous functions of $\alpha$ and exhibit a local maximum corresponding to the network topology (F). In this regime the corresponding variances, Fig.\ \ref{Fig:4}(b),  are minimal. 
This behavior provides further evidence that noise substantially modifies the basin of attraction of the individual topologies. The faster convergence rate to the topology (F) at $\alpha\in[0.001-0.003]$, together with the corresponding narrowing of the distribution of trajectories visible in Fig. \ref{Fig:3}, supports the conjecture that network self-organization into the topology (F) is a noise-induced resonance \cite{Lindner:2004}. We have verified that this behavior also occurs (i) for a relatively wide range of the input and output flows, (ii) for different exponents $n$ of the activation function, and  (iii) for a substantially larger number of demands. In general, increasing the flow leads to a larger number of redundant connections. Instead, increasing the value of the exponent $n$ in the activation function $f$ enforces the use of shortest-path connections. Interestingly, we find noise-induced phenomena for all considered values of these parameters. This also holds true when analyzing larger networks, both with respect to the grid size and the pairs of source and sink nodes, i.e., of demands (see SM \footnotemark[1]). An extensive characterization will be reported in Ref.\ \cite{Folz:2023}.

{\it Discussion.}
The noiseless equations at the basis of this study were developed in Ref.\ \cite{Tero:2007} for describing the food search of a slime mold \cite{Nakagaki:2004,Oettmeier:2020}. 
From the biological point of view, this model is oversimplified (it discards key features such as the oscillatory flow through the tubes \cite{Alim_2013,Stewart_1959}), yet it qualitatively reproduces the patterns observed in Ref. \cite{Nakagaki:2000,Tero:2010}. Moreover, it provides a powerful framework for network design and optimization algorithms \cite{Gao:2019,Li:2020}. Our work shows that the addition of noise to this model provides a qualitative improvement of the algorithmic efficiency by means of noise-induced resonances. This is a change of paradigm with regard to simulated annealing and randomized algorithms \cite{Kirkpatrick:1983,Motwani-Raghavan:1995}
and calls for a theoretical framework for stochastic nonlinear network dynamics \cite{Frank:2005,Liu:2016}.

{\it Acknowledgements.} The authors are grateful to Malte Henkel and Reza Shaebani for inspiring discussions and to Ginestra Bianconi for helpful comments. GM and FF acknowledge support from the  Deutsche  Forschungsgemeinschaft (DFG, German Research Foundation) Project-ID No.429529648, TRR 306 QuCoLiMa (Quantum Cooperativity of Light and Matter)  and from the Bundesministerium f\"ur Bildung und Forschung (BMBF, German Ministry of Education and Research) under the grant "NiQ: Noise in Quantum Algorithms". Financial support was also provided by the DFG Priority Program No. 1929 "GiRyd".
	
\bibliography{biblio-Physarum.bib} 
\end{document}


\title{Supplemental Material: noise-induced network topologies}	
	\author{Frederic Folz$^1$}
	\author{Kurt Mehlhorn$^2$}
	\author{Giovanna Morigi$^1$}
	\affiliation{$^1$Theoretische Physik, Universit\"at des Saarlandes, 66123 Saarbr\"ucken, Germany \\ $^2$Algorithms and Complexity Group, Max-Planck-Institut f\"ur Informatik, Saarland Informatics Campus, 66123 Saarbr\"ucken, Germany}
	
	\date{\today}

\maketitle

\section{Parameters and numerical simulations} 

In the model we set the length $L_{u,v}$ of the edge ${u,v}$ equal to unity when the nodes are nearest neighbors and equal to $\sqrt{2}$ when they are connected by a diagonal. We take the exponent of the activation function $n=1.2$ and set $\kappa=\gamma=1$. We impose $I_i = 0.45$ for all demands $i$. For reference, we set the potential at the node $u$ neighboring the source node of demand 1 to the right to $p_u = 0$. The initial state of the simulations has the conductivities of all edges equal to the value $D_{u, v}^0 = 0.5$. 
The conductivities are calculated by numerically integrating Eq.\ (2), together with Eq.\ (1). The calculation of the potential $p_u^i$ at node $u$ and for the demand $i$ is performed by solving a set of linear equations. Let $\ell$ be the number of network nodes and $b^i$ be the vector determining the constraint, such that it has a value of $I_i$ at the source node $s_+^i$, $-I_i$ at the sink $s_-^i$, and 0 otherwise. The vector $p^i = (p_1^i, ... p_\ell^i)$ containing all node potentials associated to demand $i$ is found by solving the linear system of equations \cite{Bonifaci:2022, Lonardi:2022}
\begin{align}
\label{Eq:A}
    M p^i = b^i\,,
\end{align}
where $M$ is a $\ell\times \ell$ matrix,  $M = ADL^{-1}A^T$ with: (i) $D = \text{diag}(D_{e_1}, ..., D_{e_m})$ the $m\times m$ diagonal matrix whose diagonal elements are the edge conductivities, (ii) $L = \text{diag}(L_{e_1}, ..., L_{e_m})$ the $m\times m$ diagonal matrix whose eigenvalues are the edge length, and (iii) $A$ the $\ell\times m$ node-arc incidence matrix of the network. In particular, the column $(A^T)_e$ has a value of 1 in position $u$ and a value of -1 in position $v$ for all edges $e = (u, v)$. 

\begin{figure}
		\includegraphics[width=0.45\textwidth]{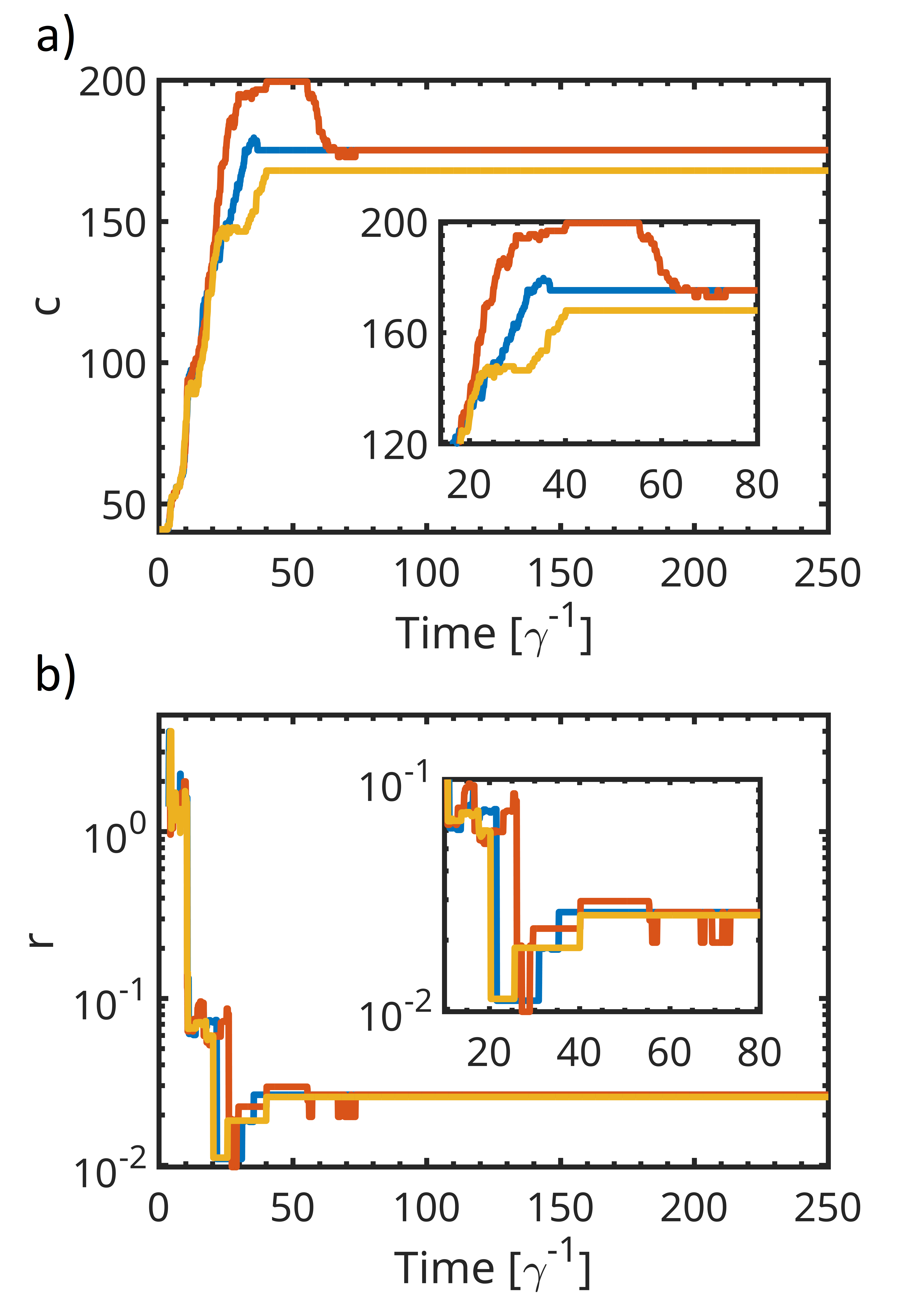}
		\caption{\label{Fig:5} The time evolution of the cost (a) and robustness (b) for three trajectories, represented by different colors. The trajectories have been determined using  $\alpha = 0.002$. The insets show a zoom in the dynamics around the convergence time $t_{\delta}^x$, see text.}
\end{figure}

The integration of Eq.\ (2) is performed using stochastic differential equations that are implemented using the Euler-Maruyama scheme with a step size of $\Delta t = 0.1 \gamma^{-1}$, as outlined in Ref.\ \cite{Kloeden_1992}. The evolution time $t_{\rm end}=250/\gamma$ is chosen after testing that each trajectory, namely, each individual evolution of the network, has reached a (meta)-stable configuration. Figures \ref{Fig:5}(a) and (b) display few trajectories at a fixed value of $\alpha$.

\section{Steady state and convergence rate } 

The reported topologies are obtained for fixed initial conditions and finite integration times, and thus, we cannot claim that they are the steady state. The steady state, in fact, is the solution of a multi-dimensional and nonlinear Fokker-Planck equation, which is not amenable of analytical treatment \cite{Frank:2005}.  Nevertheless, over the considered time, the integrated trajectories converge relatively fast towards one of the topologies of Fig.\ 2 in the main text. In order to decide whether the dynamics has converged to a (meta)stable configuration, we extend the definition of robustness and cost to a time dependent variable, which we determine on the instantaneous network's backbone, and monitor their dynamics. We identify the steady state as the configuration of the system for which the mean value is constant and the fluctuations are given by the variance set by the noise. To quantify the convergence speed, we introduce the quantity $\gamma_x$ for $x = c, r$, which has the dimensions of a rate and is defined as
\begin{equation}
\label{Eq:convergence_speed}
    \gamma_x^{-1} = \langle t_{\delta}^x \rangle\,,
\end{equation}
with $\langle \cdot \rangle$ the ensemble average over the convergence time $t_{\delta}^x$. The latter is defined as $t_{\delta}^x = \max(\{t; |x(t') - x(t_{end})| < \delta \cdot x(t_{end})\,\,\forall\,\,t' \in [t, t_{end}]\})$ with $\delta>0$. In the following, we set $\delta = 0.05$, unless otherwise stated. The rates $\gamma_x$ quantify the average convergence rate of the costs and the robustness to the steady state. Furthermore, we define the quantity $\gamma_{c, r}^{-1} = \langle \max(t_{\delta}^c, t_{\delta}^r) \rangle$, which accounts for the combined convergence time of the costs and the robustness. We calculate $\gamma_x$ for different values of the noise strength $\alpha$ by numerically solving the model given by Eqs.\ (1)-(2). Hereby, we average over 5000 simulation runs for each value of $\alpha$. In Figs.\ 4(a) and (b) in the main text the average convergence rates $\gamma_x$ and the standard deviations $\sigma_{\gamma_x}$ are shown as a function of the noise strength $\alpha$.

\section{Disparity filter} 

\begin{figure}
		\includegraphics[width=0.45\textwidth]{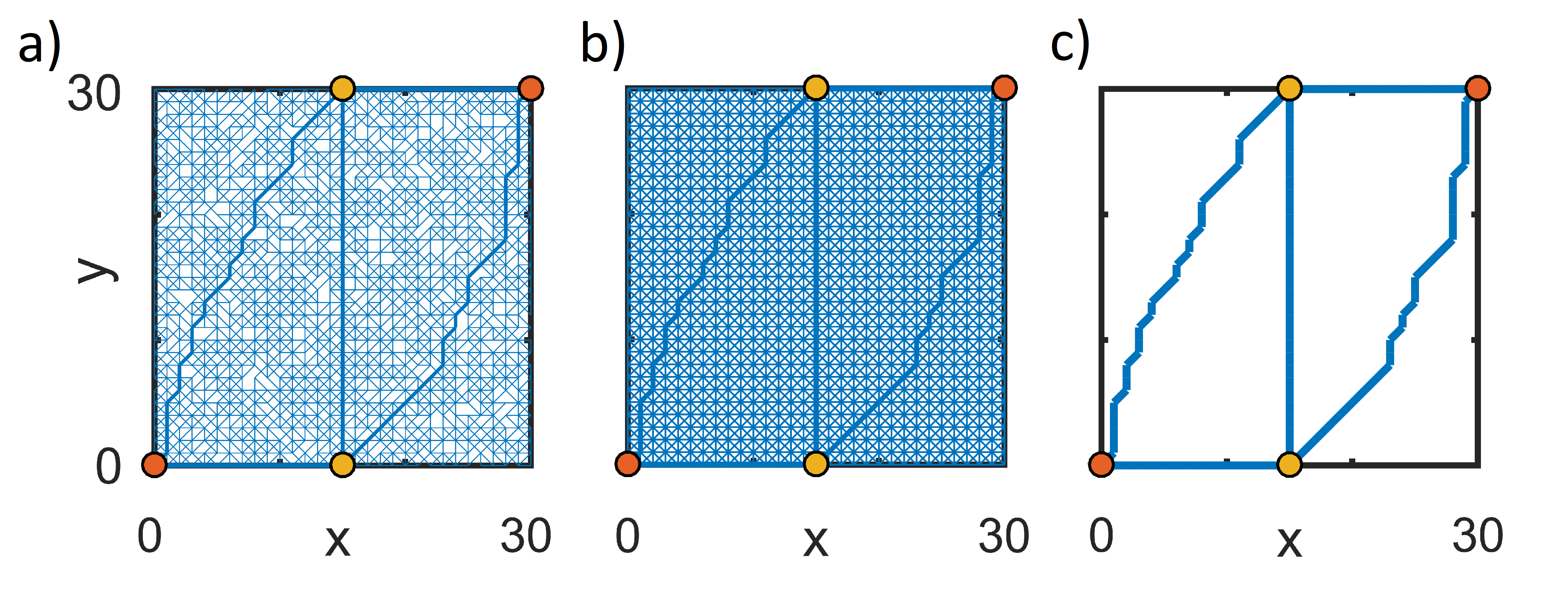}
		\caption{\label{Fig:S1} Subplot (a) displays an example of a network reached after a sufficiently long integration time in the presence of noise ($\alpha=0.002$), subplot (b) is the corresponding time average over the time interval $[249:250]/\gamma$ at the end of the simulation. The widths of the edges are graphically scaled proportionally to the corresponding conductivities. Subplot (c) displays the multi-scale backbone extracted from (b) using a filtering procedure (see text). Details on the numerical simulations are reported in the caption of Fig.\ 2 in the main text.}
\end{figure}

Extracting the network topology requires filtering the connection above a certain threshold. 
A possible ansatz consists of choosing a constant threshold for all edges. However, this approach does not account for the statistical importance that certain links of a node have over others: even if all values of the conductivities might be below threshold, some links can be statistically relevant.

In order to avoid this problem, we apply the following procedure. For each realization (trajectory) we first average the conductivities $D_{u, v}$ over the time interval $[t_{\rm end}-d_t,t_{\rm end}]$, with $d_t=1/\gamma$ corresponding to 10 time steps. The time $d_t$ is fixed by requiring that over this time the average distribution solely due to noise is stationary and is verified integrating Eq.\ ({2}) after setting $f=0$. After the averaging, the effect of fluctuations is leveled to a background value as visible by comparing Fig.\ \ref{Fig:S1}(a) with Fig.\ \ref{Fig:S1}(b), where the time averaging was performed. We then introduce a global offset $D_{u, v} \to D_{u, v} + 0.05$ for all edges $(u, v)$ and apply the disparity filter of Ref.\ \cite{Serrano:2009} using the significance level $\beta = 0.3$. The procedure of Ref.\ \cite{Serrano:2009} is implemented as follows. We determine the strength of each node $u$: $s_u = \sum_{v \in E_u} D_{u, v}$ and then normalize the conductivities of the edges that connect a node with its nearest neighbors by $p_v=D_{u,v}/s_u$ such that $\sum_{v\in E_u} p_v=1$. We remove all edges whose conductivities are not statistically significant, i.e. are purely random. As a null hypothesis, it is assumed that the edge conductivities of a certain node of degree $k$ (which can be either 8, or 5, or 3, here depending on the node location within the grid) are produced by a random assignment from a uniform distribution. In order to find the null hypothesis we use the method of induction. For $k=2$ edges we have $p_1+p_2=1$ and $p_1=x$ where $x$ is a random number in the interval $[0,1]$. We divide the interval into infinitesimal steps $dx$ and introduce the probability density $\rho(x)$ such that $p_1=\rho(x) dx$. For $k=2$, then $\rho(x)=1$. For $k>2$, we find $\rho(x)$ by solving the nested integral $\rho(x) dx= dx k\!\int_0^{1-x}dx_1\ldots\int_0^{1-x_{k-2}}dx_{k-3}$,
which gives \cite{Serrano:2009}
\begin{equation}
\label{Eq:rho}
    \rho(x) dx = (k - 1) (1 - x)^{k - 2} dx.
\end{equation}
The probability $\beta_{u, v}$ that the edge $(u, v)$ is compatible with the null hypothesis is given by
\begin{equation}
\label{Eq:disparity}
    \beta_{u, v} = 1 - (k - 1) \int_0^{\tilde D_{u, v}} (1 - x)^{k - 2} dx.
\end{equation}
The disparity filter removes all edges for which it holds $\beta_{u, v} \ge \beta$ with a significance level $\beta \in [0, 1]$ as these edges are not statistically relevant.
We note that the filter fails at sufficiently large values of the noise amplitude $\alpha$, which we do not consider here. These large values correspond to the physical situation where disorder prevails over the order imposed by the nonlinear force.

\section{Robustness of the network} 

In order to determine the robustness, we count the number of links of the filtered network. For this purpose, we assign the same conductivity to all edges of the network's backbone. We remark that various approaches to define a measure of robustness are discussed in literature. In the work of Ref.\ \cite{Tero:2010}, the fault-tolerance of a network was measured by counting the number of edges that can be removed without separating the network into two parts. Here, we chose the inverse of the total effective resistance of the network as the measure of robustness, see \cite{Ellens:2013}. This approach takes into account both the number of different paths that can be used to fulfill a demand and the paths length. Before calculating the total effective resistance, we normalize all edge conductivities $D_{u, v} > 0$ as we intend to focus on the length as the quality criterion for a path for simplicity. Extending the analysis to a measure of robustness that also takes into account the amplitude of the edge conductivities could be an interesting future consideration.

\section{Dependence on the injection current and for a larger number of demands}

\begin{figure}
		\includegraphics[width=0.45\textwidth]{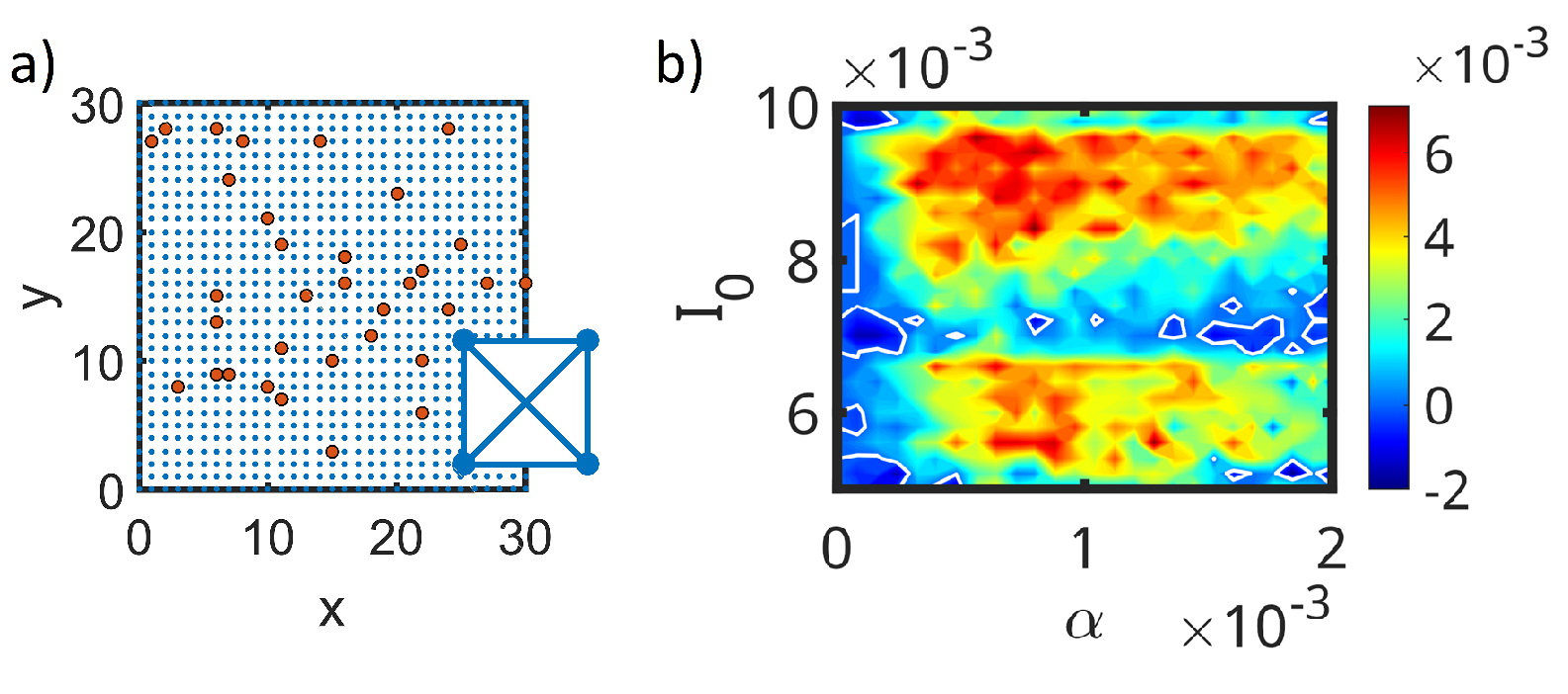}
		\caption{\label{Fig:S2} (a) Network self-organization is simulated on a grid of 31 x 31 nodes, whereby source and sink nodes are placed such that they represent the relative location of major cities around Tokyo and are labelled red. We assume that there is a demand between each pair of cities which gives rise to a total of 528 demands. The network design results from the dynamics of the edges, which are modelled by time-varying conductivity on an electrical network and in the presence of additive noise according to Eq.\ (1) and (2). The difference in robustness between the stochastic case and the deterministic case as a function of the noise and the flow $I_0$ which is the same for all demands is shown in (b). The difference in robustness is given as a color code: red means that the robustness in the stochastic case is larger than in the deterministic case and blue means the opposite. The solid white lines indicate that the robustness in the stochastic case is the same as in the deterministic case. The difference in robustness was calculated by averaging over 50 simulation runs for each pair of $I_0$ and $\alpha$.}
\end{figure}

We consider a grid with a larger number of demands in the following. In Fig.\ \ref{Fig:S2}(a), a grid of 31 x 31 nodes is shown, whereby source and sink nodes are placed such that they represent the relative location of major cities around Tokyo \cite{Tero:2010} and are labelled red. We assume that there is a demand between each pair of cities, which gives rise to a total of 528 demands. Figure\ \ref{Fig:S2}(b) displays the difference in robustness between the noiseless case and the stochastic case as a function of the noise amplitude $\alpha$ and of the flow $I_0$ (which is the same for all demands) for $n = 1.6$. The difference in robustness is given as a color code: red means that the robustness in the stochastic case is larger than in the deterministic case, blue means the opposite. The solid white lines indicate that the robustness in the stochastic case is the same as in the deterministic case. The difference in robustness was calculated by averaging over 50 trajectories for each pair of $I_0$ and $\alpha$.

In general, the noise-induced resonances appear for all values of the injection current we considered. We note that they also occur when considering different values of $I_i$ for different demands. We note that the dependence on the injected current introduces additional features, which are due to discontinuous transitions and which we will discuss elsewhere.

\section{Movies of the dynamics}

As part of the Supplemental Material, we provide movies of the dynamics leading to the networks (A)-(G) shown in Fig.\,2 of the main text:
\begin{enumerate}
    \item[-] Network (A): \url{https://www.uni-saarland.de/fileadmin/upload/lehrstuhl/morigi/noise-induced-network-topologies/network_A.gif}
    \item[-] Network (B): \url{https://www.uni-saarland.de/fileadmin/upload/lehrstuhl/morigi/noise-induced-network-topologies/network_B.gif}
    \item[-] Network (C): \url{https://www.uni-saarland.de/fileadmin/upload/lehrstuhl/morigi/noise-induced-network-topologies/network_C.gif}
    \item[-] Network (D): \url{https://www.uni-saarland.de/fileadmin/upload/lehrstuhl/morigi/noise-induced-network-topologies/network_D.gif}
    \item[-] Network (E): \url{https://www.uni-saarland.de/fileadmin/upload/lehrstuhl/morigi/noise-induced-network-topologies/network_E.gif}
    \item[-] Network (F): \url{https://www.uni-saarland.de/fileadmin/upload/lehrstuhl/morigi/noise-induced-network-topologies/network_F.gif}
    \item[-] Network (G): \url{https://www.uni-saarland.de/fileadmin/upload/lehrstuhl/morigi/noise-induced-network-topologies/network_G.gif}
\end{enumerate}
The movies cover the first 20\,\% of the total simulation time. For the networks (B, C), we used a noise amplitude of $\alpha = 0.0001$, for the network (D), we used $\alpha = 0.0008$, for the networks (E, F), we used $\alpha = 0.001$ and for the network (G), we used $\alpha = 0.005$. All other parameter values are chosen the same as described in the caption of Fig.\,2 of the main text.

\newpage

\bibliography{sm.bib}